# On the quantum nature of a fireball created in ultrarelativistic nuclear collisions

*V. A. Kizka*

*V.N. Karazin Kharkiv National University, Kharkiv, 61022, Ukraine*

*Valeriy.Kizka@karazin.ua*

**Abstract.** In the article, the fireball formed in the collision of relativistic nuclei is considered as a quantum object. Based on this, an attempt is made to explain the difference in the measurements of hyperon yields in the two experiments - NA49 and NA57. Using the basic principles of quantum mechanics, it was shown that a fireball can have two quantum states - with and without ignited Quark-Gluon Plasma (QGP). With an increase of the collision energy of heavy ions, the probability of QGP ignition increases. At the same time, the probability of the formation of fireball without QGP ignition also remains nonzero even at nuclear collision energies that are much higher than the threshold QGP formation energy, which may be erroneously considered to be fixed and which is intensively sought in modern heavy ion accelerators. Thus, at SPS energy of heavy ion collisions $\sqrt{s_{NN}}$ = 17.3 GeV, which is much higher than the assumed threshold energy of QGP formation in the region around or slightly above of $\sqrt{s_{NN}}$ = 3 GeV, only half of the central collisions of heavy ions bring to the formation of a fireball consisted of deconfined matter, the remaining half of the collisions lead to the formation of a fireball from only hadronic matter.

## 1. Introduction

The difference between the two experiments - NA49 and NA57 in the strange (hyperon) sector [1] has not yet been explained. This shows that we have missed something in understanding the nature of the fireball formed in collisions of relativistic nuclei.

A possible methodological reason for the difference in measurements between the two experiments is related to the different method for determining the centrality of the collision of two heavy ions [1], [2]. But so far this issue remains unaddressed. Moving away from this methodological issue, I consider another possible reason for the mismatch of measurements of NA49 and NA57.

The main goal of experiments on heavy ion collisions is to study the properties of Quark-Gluon Plasma (QGP). One of the main signals about the formation of QGP is an increase in the yield of hyperons, due to a decrease in the threshold energy for the formation of hyperons in comparison with that in the collision of protons [3], [4]. The almost twofold difference in the hyperon yields of the two experiments, NA49 and NA57, has not yet been discussed from the standpoint of the fundamental properties of the fireball and the processes occurring in the target when relativistic heavy ions pass through it.

The work is organized as follows. The second section shows the difference between measurements in two experiments in which a quantum object is observed. Section III shows the application of the formulas obtained in section II to the results of NA49 and NA57 in their strange sector. Section IV discusses possible methods for testing the idea discussed in the article. Section V contains conclusions.

## 2. Theoretical justification

Let us consider a quantum system described by wave function $\Psi(x)$, where *x* is a complete set of variables from which the wave function depends. Let *L* be a physical quantity (observable) that characterizes a specific property of a given quantum system. Let $L$ has discrete spectra of eigenvalues $L_a$, $L_b$ and them correspond complete set of eigenfunctions $\Psi_a(x)$ and $\Psi_b(x)$, respectively. Then we can write $\Psi(x)=a\cdot\Psi_a(x)+b\cdot\Psi_b(x)$, where $a$ and $b$ are an amplitudes of partial states $\Psi_a(x)$ and $\Psi_b(x)$, respectively. The average value of the observable $L$, multiple repeated measurements of which must be processed, is

$$\langle L \rangle = \int \Psi^*(x)\hat{L}\Psi(x)dx. \tag{1}$$

Hermitian operator $\hat{L}$ is matched to a physical value $L$. Average value $\langle L \rangle$ coincides with average value $\langle L \rangle_{\mathrm{exp}}$, obtained by statistical processing of the results of experimental measurements. Further we can write:

$$\langle L \rangle = \int \Psi^*(x)\hat{L}\Psi(x)dx = L_a a^* a \int \Psi_a^*(x)\Psi_a(x)dx + L_b b^* b \int \Psi_b^*(x)\Psi_b(x)dx = L_a a^2 + L_b b^2, \tag{2}$$

for normalized and orthogonal eigenfunctions $\Psi_a(x)$ and $\Psi_b(x)$, respectively. And $a^2$ and $b^2$ are a square of modules of amplitudes of partial states $\Psi_a(x)$ and $\Psi_b(x)$, respectively.

Suppose the first experiment *"1"* measures $L$ in an acceptance area $\Omega_1$ (kinematic and hardware conditions of the experiment that limit the signal) from which the states $\Psi_a$ and $\Psi_b$ are visible without possibility of their separation. The second experiment *"2"* make measurements in an acceptance area $\Omega_2$ from which only the state $\Psi_a$ is visible. We have the right to assume this if we assume that each of the states is characterized by its own kinematic, space-time distribution of particles emitted by the quantum system under study. (Due to the specifics of measurements on particle detectors at nuclear colliders, I have jumped ahead here, considering a quantum system as a fireball formed after a collision of nuclei). Suppose we do not know in advance which state exists at the moment of measurement. Moreover, to make matters worse, we do not even suspect that the system has two different quantum states.

The result of measurement of $L$ by experiment *"1"*: $\langle L \rangle_1 = L_a a^2 + L_b b^2$. The result of measurement of $L$ by experiment *"2"*: $\langle L \rangle_2 = L_a$, because experimental setting "2" only "sees" one state $\Psi_a$. But we assumed that we know nothing about the two states of a quantum system. This means that experiment *"2"* will measure the physical quantity $L$ even if the system is in a state $\Psi_b$. In this case, the measurement will give zero. When processing experimental data, we still take this measurement into account to calculate the average value $\langle L \rangle_2$, what will underestimate the real value in the "visible" state $\Psi_a$. Obviously, we must write the experimental value measured by experiment *"2"* in the same way as for the first experiment, but with the second term nulled: $\langle L \rangle_2 = L_a a^2$.

Let us apply to our study a good model $M_1$ that takes into account the existence of only the $\Psi_b$ state for a quantum system. The model will produce the following result for the observable $L$: $L_b^{M_1} \approx L_b$. We put an approximate equal sign, because we use a good model that gives the calculation of the observable very close to the real physical quantity. Difference between measurements of both experiments:

$$\langle L \rangle_1 - \langle L \rangle_2 = L_b b^2 \approx L_b^{M_1} b^2. \tag{3}$$

Therefore, the probability of the state $\Psi_b$:

$$b^2 = \frac{\langle L \rangle_1 - \langle L \rangle_2}{L_b^{M_1}}. \qquad (4)$$

Probability of $\Psi_a$ state is $a^2 = 1 - b^2$. Let us now apply to our study a good model $M_2$ that takes into account the existence of only the $\Psi_a$ state for a quantum system. Then:

$$\langle L \rangle_1 - \langle L \rangle_2 = L_b b^2 \neq L_a^{M_2} b^2, \qquad (5)$$

$$b^2 \neq \frac{\langle L \rangle_1 - \langle L \rangle_2}{L_a^{M_2}}, \qquad (6)$$

that is, even if both models are good, we will get a contradictory result after applying them to our study, which will suggest the existence of two quantum states in the system, the measurement of which in two different experiments also gave a contradictory result. A particular case, when the calculation results for two models coincide $L_b^{M_1} = L_a^{M_2}$, will lead to equality in (6), which, if two experiments differ, can be interpreted as the equality of two observables for two quantum states of the system $L_b = L_a$.

Formulas (4 − 6) are used in the next section for explanation of differences between NA57 and NA49 in the hyperon sector. A necessary condition for the application of these formulas is $\langle L \rangle_1 - \langle L \rangle_2 \neq 0$.

### 3. Application to experimental data

Let experiment "1" be NA57 and experiment "2" - NA49. Let the observable $L$ be the multiplicity, $L = \left. \frac{dN}{N_{wound} dy} \right|_{y=0}$, of hyperons ($\Lambda^0 + \Sigma^0$ and other hyperons) at mid-rapidity for the central $^{208}Pb + ^{208}Pb$ collisions at $\sqrt{s_{NN}} = 17.3$ GeV. $N_{wound}$ is a number of nucleons which underwent at least one inelastic collision. Or the equivalent notation $L \equiv (\Lambda^0 + \Sigma^0)/N_{wound}$ and for other hyperons. Average value measured experimentally is $\langle L \rangle \equiv \frac{1}{N_{wound} \cdot N_{ev}} \sum_{i=1}^{N_{ev}} (\Lambda^0 + \Sigma^0)_i$, where $N_{ev}$ is the number of central heavy-ion collisions selected by the trigger, $i$ is an $i$-th collision (measurement), $N_{wound}$ is fixed for central collisions of nuclei of a certain sort (in our case, these are lead nuclei $^{208}Pb$). Time interval of the fireball existence, created in central collisions, is around $15$ fm/c [5], then: $\delta t \cdot \delta E = 15$[fm/c]·$6$[MeV] = $3 \cdot 10^{-22}$ MeV·s $\approx \hbar/2$, where an uncertainty of the energy $\delta E$ was taken equal to an uncertainty of the kinetic (thermal) freeze-out temperature $T_{kin}$ from [1] for central $^{208}Pb + ^{208}Pb$ collisions obtained from the spectra of hyperons at $\sqrt{s_{NN}} = 17.3$ GeV. (An uncertainty of the kinetic (thermal) freeze-out temperature obtained by NA57 from spectra of hyperons for the most central $^{208}Pb + ^{208}Pb$ collisions is $10$ MeV at $\sqrt{s_{NN}} = 17.3$ GeV [2].) The fulfillment of the Heisenberg uncertainty relation leads us to the assumption that a fireball has some probability $p_{QGP}$ to be created with an ignition QGP inside it, and it has some probability $p_{HF}$ to be created consisting of only hadronic matter (Hadronic Fireball - HF), without QGP ignition even far beyond the probable fixed threshold energy of the QCD phase transition which is sought at modern heavy ion accelerators. In the context of the hypothesis under consideration, the assumption of fixed threshold energy should be regarded as erroneous. Probably, $p_{QGP} + p_{HF} = 1$ with both terms nonzero for any heavy ion collision energy.

We took data from both experiments NA49 and NA57 and two models - PHSD (simulates the formation of QGP, a transport model of the evolution of a fireball with parton degrees of freedom) and HSD (simulates the formation of HF, a transport model of the evolution of a fireball without parton degrees of freedom) from Fig. 21. of [5] (Table 1, the observable is $(\Lambda^0+\Sigma^0)/N_{wound}$) and substituted them into (4) (*b* is the amplitude of the QGP state, $b^2$ is $p_{QGP}$, $M_1$ is PHSD):

$$p_{QGP} = \frac{((\Lambda^0+\Sigma^0)/N_{wound})_{NA57} - ((\Lambda^0+\Sigma^0)/N_{wound})_{NA49}}{((\Lambda^0+\Sigma^0)/N_{wound})_{PHSD}}. \quad (7)$$

We have assumed here that NA57 "sees" both the states (Hadronic Fireball and QGP Fireball), and NA49 preferentially "sees" the Hadronic Fireball state. Substituting values from Table 1 (2nd column), we have the probability of creation of QGP state of matter in the central heavy-ion collisions at $\sqrt{s_{NN}}$ = 17.3 GeV is $p_{QGP} = 0.45 \pm 0.15$.

Table 1. Multiplicities of hyperons created in $^{208}$Pb+$^{208}$Pb central collisions at $\sqrt{s_{NN}}$ = 17.3 GeV at mid-rapidity measured by NA57 and NA49 Collaborations and calculated by PHSD/HSD models (taken from Fig. 21-22 of [5]).

|  | $(\Lambda^0+\Sigma^0)/N_{wound}$ | $(\bar{\Lambda}^0+\bar{\Sigma}^0)/N_{wound}$ | $(\Xi^-)/N_{wound}$ | $(\bar{\Xi}^+)/N_{wound}$ |
|---|---|---|---|---|
| NA49 | $(36.6 \pm 4.35)\cdot 10^{-3}$ | $(3.97 \pm 0.85)\cdot 10^{-3}$ | $(4.616 \pm 1.6)\cdot 10^{-3}$ | $(9.27 \pm 1.79)\cdot 10^{-4}$ |
| NA57 | $(53 \pm 3.22)\cdot 10^{-3}$ | $(6.97 \pm 0.412)\cdot 10^{-3}$ | $(6 \pm 0.2)\cdot 10^{-3}$ | $(1.46 \pm 0.1)\cdot 10^{-3}$ |
| PHSD | $36.45\cdot 10^{-3}$ | $5.85\cdot 10^{-3}$ | $3.98\cdot 10^{-3}$ | $9.27\cdot 10^{-4}$ |
| HSD | $35.85\cdot 10^{-3}$ | $2.67\cdot 10^{-3}$ | $3.16\cdot 10^{-3}$ | $2.78\cdot 10^{-4}$ |

We have to test this result for other observables and should get the same result because formulas (4 - 6) should remain the same for any observables for which we see an inexplicable difference between the measurements of the two experiments. From Fig. 21-22 of [5 the multiplicities for $\bar{\Lambda}^0+\bar{\Sigma}^0, \Xi^-$ and $\bar{\Xi}^+$ were taken for central collisions and they are shown in Table 1 (3d-5th columns). Repeating (4) for new observables, we have:

$\bar{\Lambda}^0 + \bar{\Sigma}^0 : p_{QGP} = 0.513 \pm 0.16;$

$\Xi^- : p_{QGP} = 0.34 \pm 0.4;$

$\bar{\Xi}^+ : p_{QGP} = 0.57 \pm 0.2.$

We see that all four probabilities coincide within the error limits. Averaging over these four values gives $\langle p_{QGP} \rangle$ = 0.47 ± 0.23. Thus, only half of the events of central heavy ion collisions at the considered energy ignite the QGP, the other half creates a fireball with only hadronic matter: $\langle p_{HF} \rangle$ = 1 − $\langle p_{QGP} \rangle$ = 0.53 ± 0.23.

Now suppose that NA49 "sees" only QGP fireball. Repeating the calculation for this case using (6), where we now take the HSD calculation data as the denominator (Table 1), gives the following Hadron Fireball creation probabilities:

$\Lambda^0 + \Sigma^0 : p_{HF} = 0.46 \pm 0.15;$

$\bar{\Lambda}^0 + \bar{\Sigma}^0 : p_{HF} = 1.12 \pm 0.35;$

$\Xi^- : p_{HF} = 0.43 \pm 0.5;$

$\bar{\Xi}^+ : p_{HF} = 1.9 \pm 0.72.$

We see meaningless probabilities greater than one and their large differences among themselves. Average probability $\langle p_{HF} \rangle = 0.98 \pm 0.43$. This result suggests that NA49 only "sees" the QGP state after all.

It can be assumed that the probability of creating a QGP Fireball increases with the energy $\sqrt{s_{NN}}$, which can be seen from the calculations performed for peripheral ($N_{wound} < 100$) collisions of heavy ions with the formation of hyperons, taken from Fig.21-22 of [5]: $\langle p_{QGP} \rangle = 0.23 \pm 0.26$ (calculation not shown here). Less energy pumped into the fireball resulting from peripheral heavy ion collisions is equivalent to a reduction in collision energy. This means that the chance of creating a QGP fireball increases as the energy increases.

Below is a diagram showing the formation of two fireball states after the central collision of two nuclei which have relativistic contraction before collision (on the left).

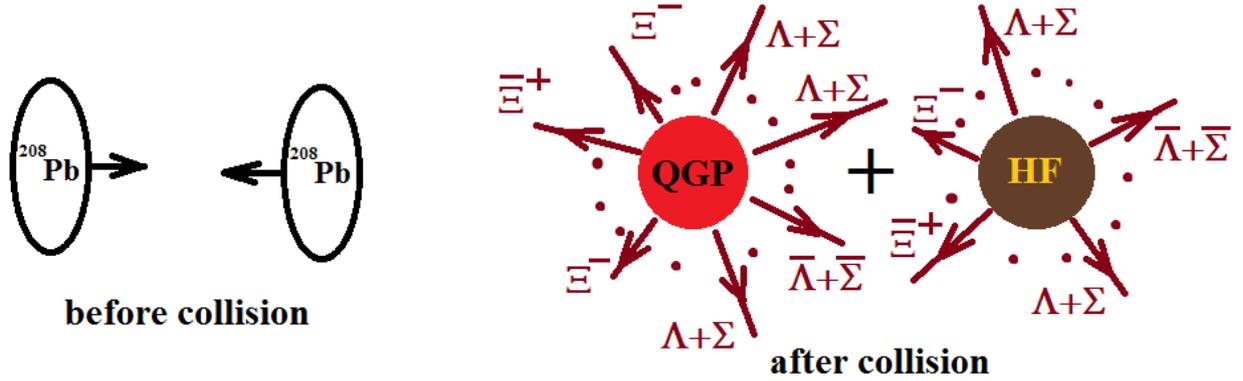

On the right, two amplitudes of two quantum phase states are shown, which cannot be experimentally separated, considering by implicit agreement, that at collision energies of heavy ions below $\sqrt{s_{NN}} = 3$ GeV, only state HF is realized [6], [7], and at energies above $\sqrt{s_{NN}} = 3$ GeV, only QGP state [8], [9]. The realization of both quantum states has never been considered, which was done in this article.

### 4. Nuances of SPS experiments

We need to consider possible reasons for the differences in measurements between the two detectors NA57 and NA49. We are dealing with the average hyperon multiplicity corresponding to $y_{CM} = 0$. Thus, we are dealing with hyperons produced in the azimuthal plane. Let experiment *1* "see" the distribution of hyperons in the interval of azimuthal angles $\Delta\phi_1$, and experiment *2* - in the interval $\Delta\phi_2$, and let $\Delta\phi_1 > \Delta\phi_2$. If there are different azimuthal dependences of hyperon production for each quantum state of the fireball, then it may turn out that experiment *2* does not "see" some extremum in the azimuthal distribution of hyperons, for example, a local maximum. But this is possible only if both experiments measure the production of hyperons in collisions of polarized nuclei, otherwise such an azimuthal dependence cannot be measured. The $^{208}Pb$ nucleus has zero magnetic moment and spin in the ground state. What about a beam of lead nuclei $^{208}Pb$?

A beam of $^{208}$Pb lead nuclei passes through two RF resonators of the SPS [10], 216 quadrupoles (max. gradient 22 T/m at a length of 3 m [11], [12]) and 744 dipole magnets (max. induction on the beam axis 2.02 T with a length of 6.5 m [11], [12] and an inevitable gradients at the edges of the magnets) of the ring of SPS before being ejected through the baffled magnets (magnetic induction 0.35 ÷ 1.5 T with length 8 ÷ 10 m [12] and inevitable gradients at the edges of the magnets) towards the target. Charged relativistic heavy ions $^{208}$Pb of the beam interact with electromagnetic fields of magnets. As a result, nuclei can be excited to levels with nonzero electromagnetic moments (dipole, quadrupole, octupole, etc.). Assuming excitation of the lead nucleus $^{208}$Pb to a state with a maximum lifetime of about 300 ps [13] and taking into account the relativistic time delay, then the ion can travel only a few meters after passing the septum magnet before de-excitation, which is not enough to reach the target, located about a hundred meters from the septum magnet [14]. Thus, an unpolarized beam of nuclei with zero spins comes to the target.

The target in both experiments is a thin lead foil $^{208}$Pb. The foil is at room temperature - the temperature of the experimental hall. It can be assumed that lead nuclei in the crystal lattice (in the crystal lattice nodes) of a thin target foil can be excited into a state with a spin and orient themselves due to electromagnetic interaction with the electric field of the crystal lattice. This interaction depends on the temperature (1), structure (2), defects (3) of the lead foil, and also on the change in the electronic and crystal structure of the solid when a relativistic beam of heavy ions moving through the crystal (4). [15] shows that the influence of the first three factors on the state of the nucleus in the crystal lattice is very small. An ultrarelativistic heavy ion $^{208}$Pb (Z = +82) loses about 7 MeV of its kinetic energy when passing through a lead foil about 100 μm thick (I calculated using the formula Bethe-Bloch). This energy is sufficient to excite the dipole or quadrupole moments of the $^{208}$Pb nucleus [13]. The beam spot area on the foil varies within 1 mm$^2$ [16]. Data on the lead foil thickness in experiments NA49 and NA57 were not found in the literature.

Then it can be assumed that the excitation of the lead nucleus (to a state with nonzero spin) of the target occurs due to electromagnetic interaction with the heavy ion incident on the target. This polarized target heavy ion collides with another heavy ion of the beam before being de-excited, forming a fireball. The polarization of an excited nucleus with a spin in some direction is determined by the structure of the crystal and the defects of the sample [15] used as a target. The interaction of polarized target nuclei with spherically symmetric beam nuclei can create an azimuthal dependence of hyperon formation.

For a preliminary assessment of the above hypothesis, it is necessary to compare the azimuthal hyperon distributions measured by two experiments, NA49 and NA57. These distributions are not presented in the published results of these experiments.

## 5. Conclusion

If we consider the fireball formed in central collisions of heavy ions $^{208}$Pb + $^{208}$Pb at $\sqrt{s_{NN}}$ = 17.3 GeV as having a quantum nature, then the probability of the formation of QGP states is about 50%. This means that about half of the collisions of heavy ions generate a fireball in which QGP ignition do not occur, which is not taken into account in any way when analyzing the experimental data. Moreover, the scientific community is firmly convinced that after reaching the threshold energy in the range around of $\sqrt{s_{NN}}$ = 3 GeV, all central collisions of heavy ions inevitably bring to the formation of a new phase of nuclear matter, as in classical physics, where, when the temperature and pressure of the phase transition are reached, the substance inevitably changes structure. The quantum nature of the fireball formed during relativistic nuclear collisions is not only ignored, no one has ever tried to consider its quantum states, each of which is characterized by own phase different from phases other quantum states that can be formed at the same collision energies but in the different events. And of course, these different quantum states, characterized by different phases, inevitably interfere with each other, which impose serious requirements on the interpretation of experimental data and on theoretical models. To date, the possibility of the coexistence of two scenarios in the evolution of hot and dense nuclear matter has not been considered by either theorists or experimenters. At this stage, the scientific community settled on the existence of one, "classical", nature of the evolution of hot and dense nuclear matter - up to the threshold energy (around $\sqrt{s_{NN}} = 3$ GeV) there is only a hadronic fireball, and after - a quark-gluon one. Of course, if more differences were found between different heavy ion experiments in measurements of the same observable, then the hypothesis presented in the article would have been pronounced long ago. But the existence of just one difference in the measurements of hyperons in two experiments, which cannot be explained for almost 15 years, forced us to consider in this article the existence of two coexisting scenarios for the evolution of the fireball.

These two scenarios imply the following important conclusion that the phase trajectory of the fireball on the QCD phase diagram should be divided into two trajectories, reflecting two possibilities for the evolution of the hot nuclear matter - with and without QGP ignition. As the heavy ion collision energy increases, the probability of QGP formation also increases, but the probability of the formation of the HF state remains nonzero even at heavy ion collision energies that are far beyond the threshold QGP formation energy.

Since both models HSD and PHSD predict mid-rapidity multiplicities of hyperon production in collisions of relativistic heavy ions of the same order of magnitude, it is impossible to impose any trigger on the events when processing experimental data. The difference in the mid-rapidity multiplicity of the production of antihyperons ($\bar{\Lambda}^0 + \bar{\Sigma}^0$ and $\bar{\Xi}^+$) by a factor of 2 or more in models predictions (HSD and PHSD) makes it possible to separate events with presumably different fireball quantum states in the processing experimental data. But this will be nothing more than an adjustment of the experiment to the proposed hypothesis. Unfortunately, the experiment NA57 was conceived only to study the strange sector, so the final resolution of the existing difference between the two experiments is possible with further operation of the NA57 facility in other sectors.